\documentclass[aps,prb,onecolumn,longbibliography,showpacs,preprint,superscriptaddress,amsmath,amssymb,floatfix,verbatim]{revtex4-2}

\usepackage{xcolor}

\usepackage{graphicx}
\usepackage{dcolumn}
\usepackage{bm}
\usepackage{xr}
\begin{document}


\title{Momentum-Resolved Relaxation-Time Approach for Size-Dependent Conductivity in Anisotropic Metallic Films}

\author{YoungJun Lee}
\affiliation{Department of Physics and Research Institute for Basic Sciences, Kyung Hee University, Seoul 02447, Korea} 

\author{Jin Soo Lee}
\affiliation{Department of Physics and Research Institute for Basic Sciences, Kyung Hee University, Seoul 02447, Korea}

\author{Seungjun Lee\footnote{Present address: Department of Electrical and Computer Engineering, University of Minnesota, Minneapolis, Minnesota 55455, USA}}
\affiliation{Department of Physics and Research Institute for Basic Sciences, Kyung Hee University, Seoul 02447, Korea}

\author{Seoung-Hun Kang}
\email[Corresponding author. email: ]{physicsksh@gmail.com}
\affiliation{Department of Physics and Research Institute for Basic Sciences, Kyung Hee University, Seoul 02447, Korea}
\affiliation{Department of Information Display, Kyung Hee University, Seoul 02447, Korea}

\author{Young-Kyun Kwon}
\email[Corresponding author. email: ]{ykkwon@khu.ac.kr}
\affiliation{Department of Physics and Research Institute for Basic Sciences, Kyung Hee University, Seoul 02447, Korea}
\affiliation{Department of Information Display, Kyung Hee University, Seoul 02447, Korea}

\date{\today} 

\begin{abstract}
Shrinking CMOS interconnect dimensions to the nanometer scale intensifies electron scattering at surfaces, interfaces, and grain boundaries, causing severe conductivity loss and challenging copper-based designs. Here we present a momentum-resolved relaxation-time framework that integrates density functional theory with the semi-classical Boltzmann transport equation to predict size-dependent resistivity in metallic thin films. Electron–phonon interactions are computed from first principles, and anisotropic surface and grain boundary scattering is captured through a momentum-dependent mean free path, allowing relaxation times to vary spatially and directionally without empirical fitting. Applied to isotropic (Cu, Ag, Au) and anisotropic (W, Ti$_2$GeC) metals, the model achieves excellent agreement with experiments and uncovers the critical role of crystallographic anisotropy in transport. We further identify layered MAX-phase compounds as promising ultrathin interconnects. This work provides a predictive, physically rigorous, and computationally efficient route to designing high-performance conductors for next-generation nanoelectronics.
\end{abstract}

\maketitle


\section{Introduction}
Copper (Cu) has long been the primary choice for interconnect applications due to its exceptional bulk conductivity and low electrical resistivity~\cite{Gupta2010}. However, as integrated circuits scale down to the nanoscale regime, copper-based interconnects face critical limitations arising from rapid increases in resistivity~\cite{Josell2009}. This significant degradation of electrical performance is predominantly driven by enhanced electron scattering on surfaces, interfaces, and grain boundaries, which fundamentally alters electron transport properties in ultrathin dimensions~\cite{Fuchs1938,Sondheimer1952,Gall2020}. Consequently, the search for alternative interconnect materials capable of mitigating these size-dependent effects has become an essential focus of recent research efforts.

To quantitatively describe the resistivity scaling phenomena in metallic thin films, many studies have employed the classical Fuchs–Sondheimer (FS) model~\cite{Fuchs1938,Sondheimer1952}, expressed as:

\begin{equation} 
\rho = \rho_{0} + \rho_{0}\lambda\frac{3(1 - p)}{4d},\label{eq1}
\end{equation}

where $\rho$ is the electrical resistivity of the thin film, $\rho_{0}$ is the bulk resistivity, $\lambda$ is the mean free path of the bulk electrons (MFP), $p$ is the specularity factor reflecting the characteristics of the surface scattering and $d$ is the thickness of the film.

Despite widespread use and simplicity, the conventional FS model exhibits two critical shortcomings. First, it relies heavily on empirical parameters, particularly the specularity factor ($p$), which encapsulates complex surface properties such as roughness, impurities, and chemical composition. Determining accurate values of $p$ requires extensive experimental characterization across various film thicknesses, which limits predictive capabilities, especially when screening new or less studied materials. Second, the FS model inherently assumes isotropic electron transport behavior, accurately describing metals with nearly spherical Fermi surfaces, such as Cu, Ag and Au; however, it fails to capture the directional electron transport inherent in anisotropic materials~\cite{Pengyuan2017}.

Many technologically relevant metals, notably tungsten (W), exhibit highly anisotropic Fermi surfaces, resulting in substantial directional variation in their electron MFP. For example, tungsten displays an MFP of approximately 18.8 nm along the [011] crystallographic direction, contrasting markedly with 33 nm along the [001] direction~\cite{Pengyuan2017}. Such a pronounced anisotropy renders conventional FS predictions inaccurate, particularly when the conduction directions and epitaxial growth orientations differ significantly. This limitation is further exacerbated in materials that possess strongly anisotropic or quasi-two-dimensional (quasi-2D) electronic structures, such as MAX-phase compounds (e.g., Ti$_2$GeC), which exhibit direction-dependent transport that cannot be captured effectively by the traditional FS model~\cite{Hadi2022,Sankaran2021}.

Density functional theory (DFT)~\cite{Kohn1965} has emerged as a powerful approach to accurately computing intrinsic electronic and phononic properties, thereby significantly improving predictive precision. However, direct application of DFT to large-scale slab geometries (tens to hundreds of angstroms), typical in interconnect research, is computationally prohibitive, rendering routine calculations impractical. Thus, there is an urgent need for computationally efficient yet quantum-mechanically rigorous models that can accurately describe anisotropic electron transport at realistic device dimensions.

In this work, we introduce and systematically validate a novel momentum-dependent relaxation-time model that accurately predicts size-dependent resistivity in metallic thin films and interconnect structures. Our methodology integrates first-principles DFT calculations of electronic structure and electron–phonon interactions with a semiclassical Boltzmann transport formalism, explicitly accounting for electron scattering at surfaces and grain boundaries without the empirical fitting parameters. By applying this approach to isotropic metals (Cu, Ag, Au), anisotropic systems (W), and strongly anisotropic MAX-phase compounds (Ti$_2$GeC), we demonstrate superior predictive performance compared to conventional FS approaches. Furthermore, our results provide critical insights into anisotropy-driven electron transport mechanisms, offering a powerful theoretical framework for discovering and optimizing next-generation interconnect materials beyond copper.

\section{\label{sec2}Method}

\subsection{Density functional theory calculations}\label{subsec0}

First-principles calculations were carried out using DFT as implemented in the Quantum Espresso code~\cite{Giannozzi2009}. The electronic wave-functions were expanded on a plane wave basis, and a suitable kinetic energy cut-off was determined through convergence tests for each material. Scalar relativistic norm-conserving pseudopotentials~\cite{Troullier1991,Schlipf2015} were used to describe the core and valence electrons, and exchange-correlation effects were treated within the Perdew-Wang~\cite{Perdew1992} local density approximation (LDA). 

The Brillouin zone was sampled using a Monkhorst-Pack $12\times12\times12$ $k$-point mesh to ensure convergence of the electronic structure. The phonon properties were then calculated using the density functional perturbation theory (DFPT)~\cite{Baroni2001} with a grid of $4\times4\times4$ $q$. Then, for each electronic state $n$ and reciprocal wave vector $k$, the electron relaxation time was evaluated as a function of energy and temperature~\cite{Noffsinger2010, Ponce2018}, focusing on the electron-phonon (el-ph) scattering channel, which dominates at room temperature according to the Fermi liquid theory~\cite{Echenique2000}. 

To achieve a finer sampling of the Brillouin zone, maximally localized Wannier functions (MLWFs)~\cite{Marzari1997} were constructed using the EPW package~\cite{Noffsinger2010}, enabling interpolation onto a $60\times60\times60$ $k$-point mesh and random sampling of 10,000 $q$-points. From these MLWFs, the relaxation time was further refined using EPW, which incorporates electron-phonon coupling. Finally, this relaxation time was incorporated into the Boltzmann transport equation using the BoltzTrap package~\cite{Madsen2006}, allowing the evaluation of the electrical conductivity of each system.

\subsection{The size-dependent resistivity model}\label{sec3}

\subsubsection{Resistivity of bulk}\label{subsec1}

The electron relaxation time in a bulk system is generally described by the Matthiessen’s rule~\cite{Dugdale1967}, which accounts for multiple scattering mechanisms:

\begin{eqnarray} 
\frac{1}{\tau} = \frac{1}{\tau_{\mathrm{el-el}}}+\frac{1}{\tau_\mathrm{el-ph}}+ \frac{1}{\tau_\mathrm{GB}}+\cdots \approx \frac{1}{\tau_\mathrm{el-ph}} 
\label{eq2} 
\end{eqnarray}

where $\tau_\mathrm{el-el}$ is the relaxation time for electron-electron scattering and $\tau_\mathrm{el-ph}$ is the relaxation time for el-ph scattering~\cite{Noffsinger2010}. According to Fermi liquid theory, el-el scattering rates follow a parabolic dependence on energy, reaching a minimum at the Fermi level~\cite{Echenique2000}. At room temperature, the el-el scattering rate is considerably lower than that of el-ph scattering. Hence, under the relaxation time approximation, $1/\tau_\mathrm{el-el}$ can be safely neglected in Eq.~(\ref{eq2}). 

Furthermore, $\tau_\mathrm{GB}$ represents the relaxation time associated with grain boundary and defect scattering. If defects and grain boundaries are sufficiently sparse and contribute only minor scattering compared to phonons, the total relaxation time can be approximated by $\tau_\mathrm{el-ph}$ alone, as shown in Eq.~(\ref{eq2}).

To determine the el-ph relaxation time more rigorously, we compute the energy- and temperature-dependent relaxation time $\tau(E_{n,\textbf{k}}, T)$ from the imaginary part of the electron self-energy~\cite{Feliciano2007}:

\begin{equation}
{\begin{aligned}
\frac{1}{\tau(E_{n,\mathbf{k}},T)} &= \pi \sum_{m,\nu} \int \frac{d\mathbf{q}}{\Omega_{BZ}} |\mathcal{G}_{mn,\nu}(\mathbf{k},\mathbf{q})|^{2} \\ 
&\times \Bigg\{ 
    \left[ f(E_{n,\mathbf{k}+\mathbf{q}}) + g(\omega_{\nu,\mathbf{q}},T) \right] 
    \delta(E_{n,\mathbf{k}} - E_{m,\mathbf{k}+\mathbf{q}} + \omega_{\nu,\mathbf{q}}) \\
&\quad + \left[ 1 - f(E_{n,\mathbf{k}+\mathbf{q}}) + g(\omega_{\nu,\mathbf{q}},T) \right] 
    \delta(E_{n,\mathbf{k}} - E_{m,\mathbf{k}+\mathbf{q}} - \omega_{\nu,\mathbf{q}}) 
\Bigg\}
\end{aligned}}
\label{eq3}
\end{equation}

where $\Omega_{BZ}$ is the Brillouin zone volume,
$\mathcal{G}_{mn,\nu}(\textbf{k},\textbf{q})$ is the el-ph matrix element, and
$f(E_{n,\textbf{k}+\textbf{q}})$ and $g(\omega_{\nu,\textbf{q}},T)$ are the Fermi-Dirac and Bose-Einstein distribution functions, respectively. Here, $\delta$ denotes the Dirac delta function, $E$ is the electron energy, $\omega$ is the frequency of phonons and $T$ is the temperature of the system.

At room temperature (300K), we use the computed relaxation time $\tau(n,\textbf{k})$ to evaluate the bulk electrical conductivity tensor via the semiclassical Boltzmann transport equation:

\begin{eqnarray} \sigma_{\alpha \beta} = \frac{2 e^{2}}{8\pi^3 \hslash} \sum_{n} \int\int_{S_F^n} {\frac{\tau(n,k) v_{\alpha}(n,k) v_{\beta}(n,k)}{\left| v(n,k)\right|} d\textbf{S}} \label{Boltzmann transport} \end{eqnarray}

where $\alpha, \beta$ are Cartesian indices (x, y, z), $e$ is the elementary charge and $\hslash$ is the reduced Planck constant. The term $\sum_{n} \int\int_{S_F^n} d\textbf{S}$ indicates surface integration over the $n$th Fermi surface, and $v(n,k)$ is the group velocity of the electron.

This formulation incorporates size-dependent resistivity by capturing both anisotropic electronic properties and el-ph interactions in the transport model. By employing this approach, we aim to establish a more precise predictive framework for resistivity scaling in ultra-thin metallic interconnects.

\subsubsection{Momentum-dependent MFP model}\label{subsec2}

\begin{figure}
\includegraphics[width=1.0\textwidth]{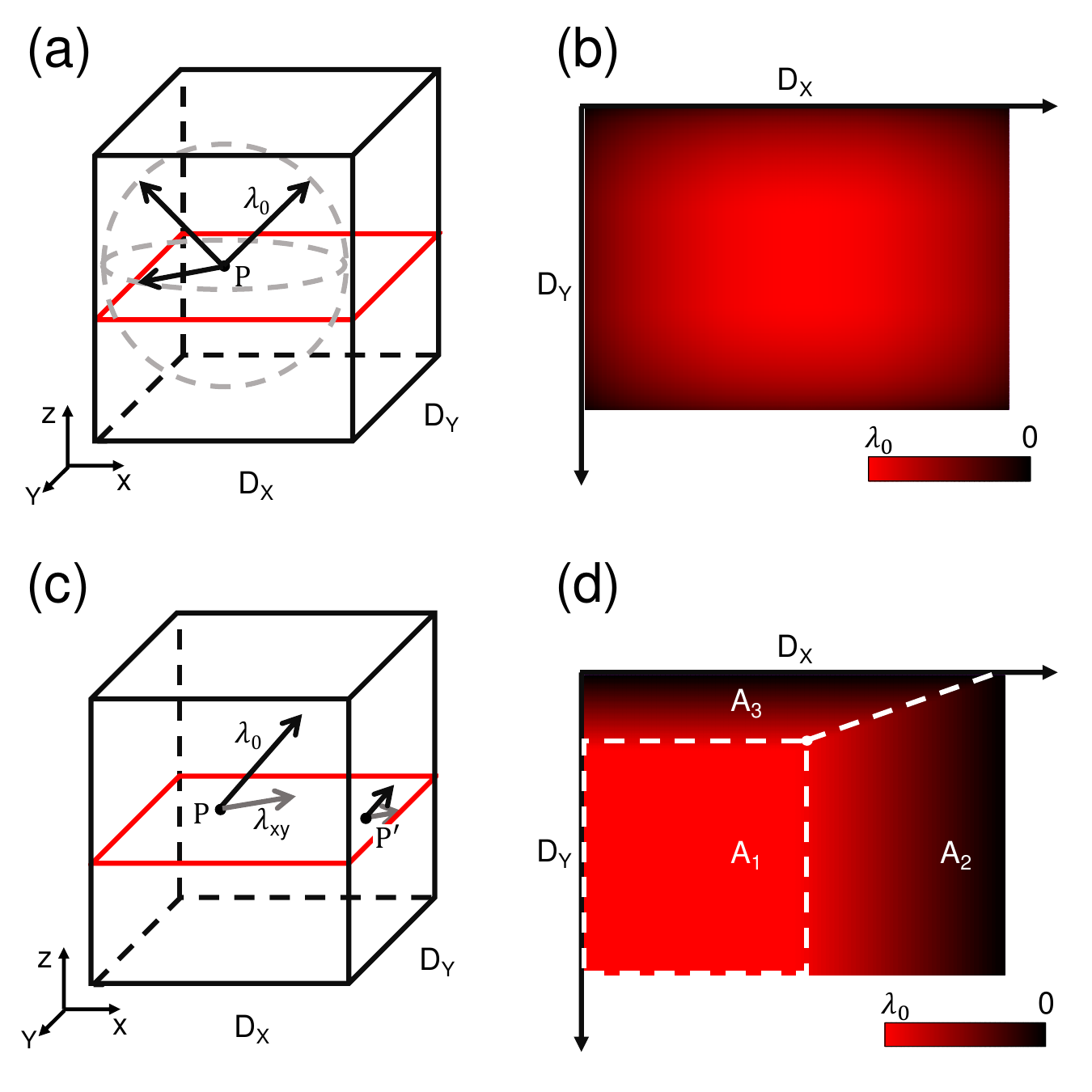}
\caption{Schematic representation of the mean free path (MFP) distribution in a rectangular wire conductor. (a) Illustration of MFP propagation from an arbitrary point P based on the Thomson random propagation model, assuming isotropic electron scattering. (b) Modified MFP distribution within the conductor cross-section under the Thomson random propagation model, demonstrating the effects of surface scattering. (c) Directional MFP propagation under the momentum-dependent MFP model, incorporating anisotropic transport effects.(d) Modified MFP distribution in the conductor cross-section derived from the momentum-dependent MFP model. In this approach, region A$_1$ remains largely unaffected by surface scattering, preserving its bulk MFP, while regions A$_2$ and A$_3$ exhibit a reduction in MFP due to interactions with the surface boundaries.}\label{fig1}
\end{figure}

In this section, we introduce a novel method for evaluating the resistivity of thin-film conductors by extending Thomson's random propagation model under modified assumptions. Consider a scenario in which the wire thickness becomes smaller than the bulk MFP ($\lambda_0$), causing surface scattering to significantly affect electron transport. Thomson's original model~\cite{Thomson1901} is based on two primary assumptions:

1. "Directional isotropy": Electrons scatter at any point on the wire with equal probability, regardless of their initial or final directions.

2. "Surface-scattering criterion": Scattering occurs only when an electron encounters the surface, which modifies its MFP accordingly. 

Figure~\ref{fig1}(a) illustrates the application of Thomson's random propagation model to a rectangular wire conductor. Based on the two assumptions above, the expected MFP distribution across the cross-sectional area of the conductor appears in Fig.~\ref{fig1}(b). Near the surface, the MFP is reduced by increased scattering. However, most real materials have anisotropic Fermi surfaces, which means that the Fermi velocity varies in both magnitude and direction for different $k$-points and electronic bands.

As a result, any model that adopts Thomson's first assumption (isotropic electron propagation) inherently neglects the momentum-dependent transport effect. This limitation becomes critical for materials exhibiting strong directional dependencies, where specific electron momentum states significantly alter the scattering behavior. To address this, we must modify Thomson's model to incorporate momentum-dependent transport, accurately reflecting the directional nature of electron scattering in real materials.

\begin{figure}
\includegraphics[width=1.0\textwidth]{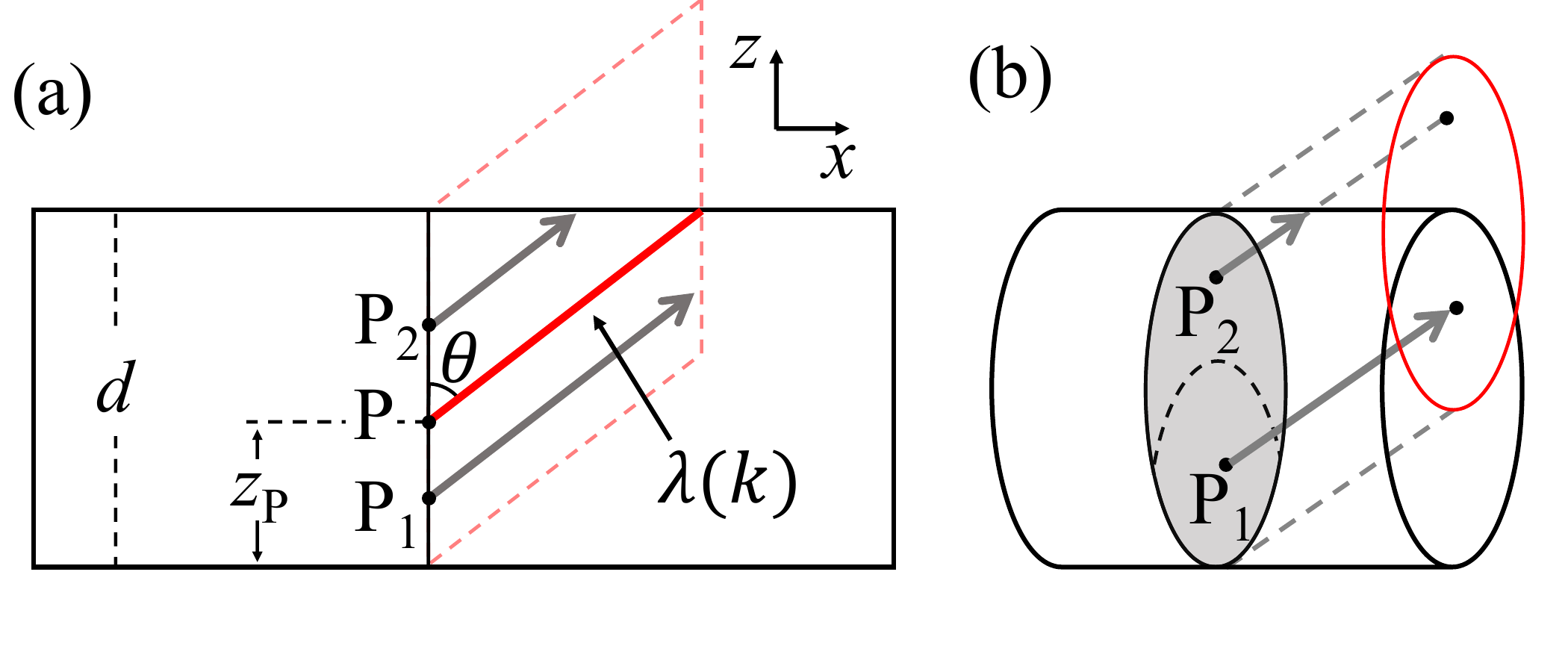}
\caption{Direction-dependent modified mean free path (MFP) models. (a) The thin-film model is confined in the $z$-direction while being infinite in the $xy$-plane. (b) The cylinder-wire model extends infinitely in the $x$-direction only. The angle $\theta$ represents the inclination between the thin-film surface and the group velocity vector for a given band index $n$ and wave-vector $k$. The arrows originating from points P$_1$ and P$_2$ illustrate the MFP that electrons can traverse from each boundary location.}\label{fig2}
\end{figure}
To extend Thomson’s model beyond isotropic systems, we relax the second assumption to include momentum-dependent transport for a given band index ($n$) and wave vector ($\mathbf{k}$). In this approach, the group velocity ($\mathbf{v}(n,\mathbf{k})$) determines the direction of electron transport rather than just the spatial position. This modification allows us to describe both isotropic and anisotropic systems with more accuracy. 
The electron group velocity is defined as:
\begin{equation}
\mathbf{v}(n,\mathbf{k})=\frac{1}{\hbar}\nabla_{\mathbf{k}}E(n,\mathbf{k}) \\
\label{eq:v_nk}
\end{equation}
where ${\mathbf{k}}E(n,\mathbf{k})$ is the electronic band structure and $\hbar$ is the reduced Planck constant.

Because MFP ($\lambda$) depends on $\mathbf{v}(n,\mathbf{k})$, it also contains direction information:
\begin{equation}
\boldsymbol{\lambda}_{n,\mathbf{k}}=\tau(n,\mathbf{k})\mathbf{v}(n,\mathbf{k})
\label{eq:lambda_nk}
\end{equation}

We apply this directional MFP concept to a simplified thin-film geometry [Fig.~\ref{fig2} (a)], where the film is infinite in the $xy$-plane and finite in the $z$-direction. If an electron originates from P$_1$, it can retain its bulk MFP ($\lambda_0$) (no surface scattering). In contrast, an electron starting from point P$_2$ at an angle ($\theta$) to the surface will experience surface scattering, reducing its MFP. The reduction continues until the distance traveled in the $z$-direction satisfies:
\begin{equation} \lambda(k) \cos{\theta} = z. \label{eq:theta}\end{equation}

Thus, the size-dependent MFP $\boldsymbol{\lambda}_{d,z}(n,\mathbf{k})$ is given by:
\begin{eqnarray}
\lambda_{d,z}(n,\mathbf{k}) =
\begin{cases}
\lambda_0(n,\mathbf{k})  &   ( 0 \le z \le z_\mathrm{P} ) \\
\frac{z}{\cos{\theta(n,\mathbf{k})}} , & ( z_\mathrm{P} < z \le d )
\end{cases}\label{eq:lamdbadz}
\end{eqnarray}
where $d$ is the total thickness of the film and $z_P$ is the critical depth at which surface scattering begins. Assuming the electronic structure of the film remains bulky, the Fermi velocity (and therefore the group velocity) remains unchanged. Consequently, the modified relaxation time becomes:
\begin{eqnarray}
\tau_{d,z}(n,\mathbf{k}) =
\begin{cases}
\tau_0(n,\mathbf{k})  &   ( 0 \le z \le z_\mathrm{P} ) \\
\frac{z}{\cos{\theta(n,\mathbf{k})} v(n,\mathbf{k})} , & ( z_\mathrm{P} < z \le d )
\end{cases}\label{taudz}
\end{eqnarray} 

The expectation value of the relaxation time is:
\begin{equation}
\tau_{d}(n,\mathbf{k}) = \frac{1}{d}\Big\{\int^{z_\mathrm{P}}_{0}{\tau_0(n,k)dz} + \int^{d}_{z_\mathrm{P}}{\frac{z}{\cos{\theta(n,\mathbf{k})} v(n,\mathbf{k})}}dz\Big\}
\label{eq:tau_d}
\end{equation} 

To include this momentum-dependent relaxation time in transport calculations, we substitute $\tau(n,\mathbf{k})$ into the semi-classical Boltzmann transport equation: 
\begin{eqnarray}
\sigma_{\alpha \beta} = \frac{2 e^{2}}{8\pi^3 \hslash} \sum_{n} \int\int_{S_F^n} {\frac{\tau_{d}(n,\mathbf{k})v_{\alpha}(n,\mathbf{k}) v_{\beta}(n,\mathbf{k})}{|v(n,\mathbf{k})|}  dS} 
\label{eq:modi boltzmann transport film}
\end{eqnarray}

thereby capturing the direction-dependent impact of surface scattering. If the effect of MFP in the $z$-direction becomes smaller than the thickness of the film $d$, $z_{P}$ reduces to 0 in Eq.~(\ref{eq:lamdbadz}). As $d$ increases, electrons cannot reach the boundary without scattering, lowering the MFP and thus increasing the resistivity in thinner films, an important effect in interconnect scaling.

\subsubsection{Momentum-dependent MFP model in a wire model}\label{subsec3}
In Sect.~\ref{subsec2}, we introduced Eqs.(\ref{eq:tau_d}) and (\ref{eq:modi boltzmann transport film}), which define a modified relaxation time and the corresponding Boltzmann transport equation for the thin-film model. We now generalize this framework to a wire model with arbitrary boundary conditions, as shown in Fig.~\ref{fig2}(b).

In the wire model, the region near the point P$_1$, denoted S$_1$, remains unaffected by boundary conditions and preserves its bulk MFP. In contrast, the region containing P$_2$, denoted S$_2$, experiences boundary-induced modifications to its relaxation time. The extent of these changes depends on various factors, including the shape of the wire boundary, the angle between the MFP and the conduction direction, and the position of $P$.

Let $\lambda(p,r,\theta)$ represent the MFP within region S$_2$, where boundary effects are significant. The overall expectation value of the relaxation time can then be expressed as:
\begin{equation}
\tau_{d}(n,\mathbf{k}) = \frac{\int_{S_1}{\tau_0(n,\mathbf{k})d\mathbf{S}} + \int_{S_2}{\lambda(p,r,\theta)/ v(n,\mathbf{k}) d\mathbf{S}}}{\int_{S_1+S_2} d\mathbf{S}}
\label{tau_gen}
\end{equation} 

providing a weighted average of the relaxation time across the wire cross-section. In Fig.~\ref{fig2}(b), the cross-sectional region containing $P$ is shown in gray, and the red circle depicts an area generated by displacing the gray region along the local MFP. Dividing the resulting volume by $\lambda_0(n,\mathbf{k}) v(n,\mathbf{k})$ leads to Eq.~(\ref{tau_gen}), capturing the boundary-scattering effects on the relaxation time.

Finally, substituting $\tau_{d}(n,k)$ from Eq.~(\ref{tau_gen}) into Eq.~(\ref{eq:modi boltzmann transport film}), we obtain the conductivity tensor for the wire model under general boundary conditions. This extension supports a more comprehensive treatment of electron transport, accommodating complex geometries and anisotropic boundary effects. 

\subsubsection{Modified MFP in a rectangular wire}\label{subsec4}
The scattering behavior at wire boundaries highlights the key difference between the momentum-dependent MFP model and the Thomson’s random propagation model, which assumes isotropic electron propagation with equal probability in all directions. Consider a rectangular wire extending infinitely along the $z$-axis, with thickness $\mathrm{D}_{x}$ and width $\mathrm{D}_{y}$, where conduction proceeds along the $z$-direction. At a given $(n,\mathbf{k})$, the bulk MFP $\lambda_{0}$ from point P (black arrow) projects onto the $xy$-plane as $\lambda_{xy}$. In the momentum-dependent MFP model, $\lambda_{xy}$ determines the modified MFP based on the position of point P, following an approach similar to Eq.(\ref{eq:lamdbadz}). Unlike Thomson’s model, which does not consider directional dependence, this method produces a more accurate assessment of boundary scattering effects.

As shown in Fig.~\ref{fig1}(a) the cross section is partitioned into three regions: $\mathrm{A}_{1}$, $\mathrm{A}_{2}$, and $\mathrm{A}_{3}$, based on $\lambda_{xy}$. Region $\mathrm{A}_{1}$ (Fig.~\ref{fig1}(b)) corresponds to areas where the change $\lambda_{xy}$ does not reduce the modified MFP - analogously to the region $\mathrm{S}{1}$ in Sect.~\ref{subsec2}- whereas $\mathrm{A}_{2}$ and $\mathrm{A}_{3}$ exhibit a linear decrease in the modified MFP as the boundary is approached, due to surface scattering. Averaging over the entire cross-section yields the effective modified MFP, $\lambda_{\textbf{D}_{x},\textbf{D}_{y}}(n,\mathbf{k})$. For isotropic electron propagation, the modified MFP is obtained by averaging the paths depicted in Fig.~\ref{fig1}(c). For example, with $\mathrm{D}_{x} = 10$ nm, $\mathrm{D}_{y} = 12$ nm, and $\lambda_{0} = 20$ nm, Thomson’s model predicts $\lambda_{\mathrm{D}_{x},\mathrm{D}_{y}}(n,\mathbf{k}) = 11$ nm, while the momentum-dependent MFP model gives $\lambda_{\mathrm{D}_{x},\mathrm{D}_{y}}(n,\mathbf{k}) = 13$ nm. 

When the Fermi surface is isotropic and all MFPs at the Fermi level are equivalent, both models coincide. However, as anisotropy increases, the discrepancy between the two approaches grows. The degree of deviation depends on the anisotropic electronic structure in the material, the geometry of the conductor, and the spatial variation of the MFPs. This comparison highlights the limitations of Thomson's model, which ignores momentum dependence, and shows that the momentum-dependent MFP model provides a more accurate representation of size-dependent resistivity, particularly in materials with strongly anisotropic electronic properties.

\subsubsection{Size-dependent resistivity in a film model with grain boundaries}\label{subsec5}

In conductive materials, electron scattering at grain boundaries significantly influences electrical transport properties, particularly in thin films. Although bulk metals typically exhibit grain sizes large enough to minimize grain boundary scattering effects, thin films rarely achieve comparable grain sizes due to growth constraints~\cite{Smith1948}. Two main factors limit grain growth in thin films: (i) strong substrate interactions restrict lateral grain expansion, restricting grains from freely developing; and (ii) geometric confinement imposed by surface and interface boundaries further inhibits grain boundary migration and grain growth~\cite{Dulmaa2021}. To systematically quantify the impact of grain size limitations in thin films, we modeled the grain growth dynamics of cobalt films as a function of thickness, allowing us to accurately simulate grain boundary spacing, $L(d)$, and subsequently analyze the associated contributions of grain boundary scattering to resistivity~\cite{Kozlowski2017}.
Calculating the relaxation time directly for a bulk system containing grain boundaries can be challenging because of instability in the calculations. To simplify this problem, we employ an alternative model that captures the essential physics in a more tractable form. First, consider two bulk systems with translation symmetry connected by a grain boundary, whose total resistivity $\rho$ is given by

\begin{eqnarray}
\rho=\rho_{0}+\frac{\gamma_\mathrm{GB}}{L_\mathrm{GB}}
\label{eq:rho_GB}
\end{eqnarray}

where $\rho_{0}$ is the base resistivity of the bulk material, $\gamma_\mathrm{GB}$ is the additional contribution of the resistivity from the grain boundary and $L_\mathrm{GB}$ is the average distance between the grain boundaries.
Next, we extend this approach to a thin-film system of thickness $d$ in the presence of grain boundaries:

\begin{eqnarray}
\rho(d)=\rho_{0}(d)+\frac{\gamma_\mathrm{GB}}{L(d)}
\label{eq:rho_GB film}
\end{eqnarray}
\begin{eqnarray}
L(d)=
\begin{cases}
L_\mathrm{GB}  &   ( L_\mathrm{GB} \le d ) \\
de^{1-\frac{d}{L_\mathrm{GB}}} & ( d < L_\mathrm{GB} )
\end{cases}
\label{eq:effective L_GB}
\end{eqnarray} 

where $d$ is the film thickness, and $L(d)$ is an effective grain-boundary spacing that accounts for the reduction in grain size as the film thickness decreases due to growth constraints imposed by the film surface.
Even though the out-of-plane direction of the film is perpendicular to the grain boundary, changes in film thickness still affect electron transport by altering the available cross-sectional area, thus influencing the overall resistivity.

1. **Case: $d \geq L_\mathrm{GB}$**
   When the film thickness is larger than the average grain-boundary spacing, electrons encounter grain boundaries with roughly the same probability as in the bulk. Hence, $\rho(d)\approx\rho_{0}(d) +\gamma_\mathrm{GB}/L_\mathrm{GB}$

2. **Case: $d < L_\mathrm{GB}$**
   When the film thickness decreases below the typical grain boundary spacing observed in bulk materials, substrate and interface interactions significantly limit grain growth, resulting in reduced effective grain sizes. Consequently, the effective grain boundary spacing, $L(d)$, decreases at smaller thicknesses, increasing the probability of electron scattering at grain boundaries. Given that the grain boundary resistivity parameter ($\gamma_\mathrm{GB}$) is constant for a given material, this reduced $L(d)$ directly contributes to an elevated total resistivity, highlighting the critical influence of thickness-dependent grain structures on electron transport in ultra-thin conductive films.
   
Moreover, if $L_\mathrm{GB}$ becomes extremely small, the resistivity of the system can reach very large values. 
As a result, materials with large grain boundary sizes in bulk exhibit higher resistivity due to grain boundary scattering.
Conversely, when $d = L_\mathrm{GB}$, the model should remain continuous and differentiable, reflecting a smooth transition between these regimes. Equation~(\ref{eq:effective L_GB}) follows from enforcing these continuity and differentiability conditions.

Finally, by fitting the two parameters $\gamma_\mathrm{GB}$ and $L_\mathrm{GB}$ to experimental or computational data, we can infer the grain-boundary purity and structural properties of a specific system. In this way, the model provides a practical means of capturing grain-boundary effects on size-dependent resistivity in thin films.

\section{Result}\label{sec4}

\subsection{Momentum-Dependent Mean-Free-Path Model for Thin Films of Isotropic Metal}\label{subsec6}

\begin{figure}
\includegraphics[width=1.0\textwidth]{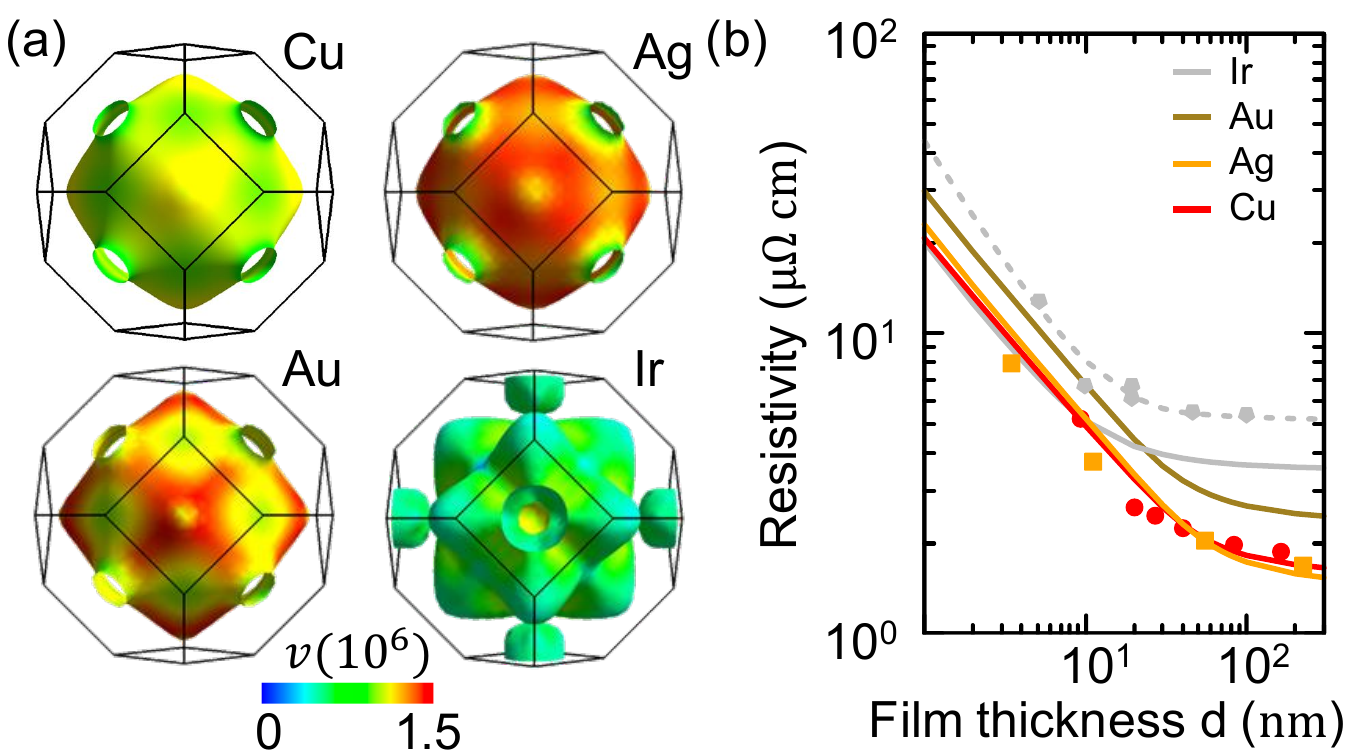}
\caption{ The Fermi surface and the thickness-dependent resistivity for isotropic metals(Cu, Ag, Au, and Ir). (a) Calculated Fermi surfaces of representative isotropic metals, exhibiting nearly spherical geometries characteristic of their isotropic electronic structures. (b) Comparison of experimental resistivity measurements (symbols) and theoretical predictions (solid lines) for Cu, Ag, Au, and Ir thin films. Model predictions incorporate momentum-dependent mean-free-path (MFP) calculations, highlighting the accuracy of our approach. The dashed lines indicate resistivity curves that include additional grain boundary scattering effects.}
\label{fig3}
\end{figure}
To validate our momentum-dependent mean free path (MFP) model, we applied it to a series of well-characterized noble metals known for their isotropic electron transport. Figure~\ref{fig3}(a) illustrates the typical isotropic behavior of Cu, Ag, Au, and Ir—each featuring monatomic composition and nearly spherical Fermi surfaces. Although traditional models already reliably predict resistivity for these systems, our goal was to investigate whether incorporating momentum directionality improves predictive accuracy for size-dependent resistivity in thin metallic films. 

We started our study with a first-principles evaluation of the resistivity using the electron-phonon(el-ph) relaxation time at 300 K using Eq.~\ref{eq3} for isotropic metals such as Cu, Ag, Au, Ir, detailed for Cu and Ir in the Supplementary Information Fig.S1 and Fig.S2. By interpolating relaxation times on dense momentum grids, we obtained a theoretical resistivity of 1.59 $\mu\Omega\cdot\mathrm{cm}$ for Cu, which closely matches the experimental value of 1.68 $\mu\Omega\cdot\mathrm{cm}$~\cite{serway1998}.

Subsequently, the predictions of our model were systematically compared with the experimental resistivity for other noble metals. For Au, the theoretical resistivity of 2.38 $\mu\Omega\cdot\mathrm{cm}$ aligns closely with the experimental value of 2.44 $\mu\Omega\cdot\mathrm{cm}$~\cite{serway1998,matula1979}. Similarly, Ag shows good agreement, with a calculated value of 1.46 $\mu\Omega\cdot\mathrm{cm}$ versus the experimental 1.59 $\mu\Omega\cdot\mathrm{cm}$~\cite{serway1998}. For Ir, the theoretical resistivity is 3.51 $\mu\Omega\cdot\mathrm{cm}$, somewhat below the measured 4.71 $\mu\Omega\cdot\mathrm{cm}$~\cite{serway1998}.

This general consistency between the calculated and measured resistivity underscores the reliability of our computational approach. However, notable deviations emerge for Ir as compared to those of other noble metals. Although Cu, Ag, Au, and Ir all possess face-centered cubic (FCC) structures, differences in grain boundary energy result in distinct grain growth characteristics. Specifically, Cu, Ag, and Au exhibit rapid grain growth rates as a result of low grain boundary energies. In contrast, Ir exhibits notably smaller grains due to its higher grain boundary energy~\cite{chen1992}. Consequently, theoretical resistivity calculated under the assumption of exclusive electron–phonon scattering significantly underestimates experimental resistivity values, where grain boundary scattering inevitably contributes. A detailed analysis of the role of grain boundary scattering is provided in Sect.~\ref{subsec8}. The current section focuses exclusively on intrinsic size-dependent resistivity behavior, independent of grain boundary effects.

The application of our momentum-dependent MFP model to thin films further highlights its predictive capabilities, as illustrated in Fig.~\ref{fig3}(b). The model accurately reproduces the size-dependent resistivity trends for all isotropic metals studied, confirming its broader applicability across diverse metallic systems.
For thicker films, the calculated resistivity closely matches the established bulk values. Given the strong agreement between bulk electron-phonon scattering-only resistivity and experimental measurements, it is expected that the grain boundary ($\rho_\mathrm{GB}$) and defect scattering ($\rho_\mathrm{defect}$) have minimal influence on resistivity. Therefore, Fig.~\ref{fig3}(b) demonstrates a strong consistency between experimental thin-film resistivity and our model predictions even at reduced film thicknesses.
Among these metals, Cu and Ag possess particularly low intrinsic resistivities; yet, because of their large bulk MFP values, resistivity increases span broader thickness ranges, resulting in steeper slopes compared to those of Au and Ir. Notably, Ir exhibits a delayed resistivity increase, sharply rising when the film thickness approaches its characteristic bulk MFP. This behavior is quantitatively captured by Eq.~\ref{eq:lamdbadz}, highlighting how the critical depth $z_p$ is intrinsically tied to the bulk MFP.
For films thicker than bulk MFP ($d > \lambda_\mathrm{bulk}$), significant contributions from intrinsic bulk MFP, $\boldsymbol{\lambda}_0(n,\mathbf{k})$, persist within the region $0 \leq z \leq z_p$, the moderating resistivity increases. In contrast, for thinner films ($d \leq \lambda_\mathrm{bulk}$), the critical depth abruptly shrinks to zero, drastically reducing the effective size-dependent MFP, $\boldsymbol{\lambda}_{d,z}(n,\mathbf{k})$, and thus causing rapid increase in resistivity.
This characteristic is clearly discernible in the resistivity profiles shown. For example, Cu and Ag, both characterized by a bulk MFP around 39 nm~\cite{haynes2016}, display resistivity slope transitions near this critical thickness. In contrast, Ir, with its significantly shorter bulk MFP of approximately 6.4 nm~\cite{haynes2016}, exhibits pronounced resistivity increases at notably thinner dimensions.
These observations firmly validate our momentum-dependent MFP approach, demonstrating its capability to accurately correlate resistivity enhancements to the interplay between thin-film thickness and intrinsic bulk electronic transport characteristics.
Furthermore, as shown in Supplementary Information Fig.S3 and Fig.S4, copper films grown along the (111) and (110) directions exhibit nearly identical relaxation time distributions and resistivity trends as a function of film thickness. This consistency further confirms the isotropic nature of copper, in which crystallographic orientation has a negligible impact on electron transport properties.

\subsection{Momentum-Dependent Mean-Free-Path Model for Thin Films of Anisotropic Metal}\label{subsec7}

\begin{figure*}
\includegraphics[width=1.0\textwidth]{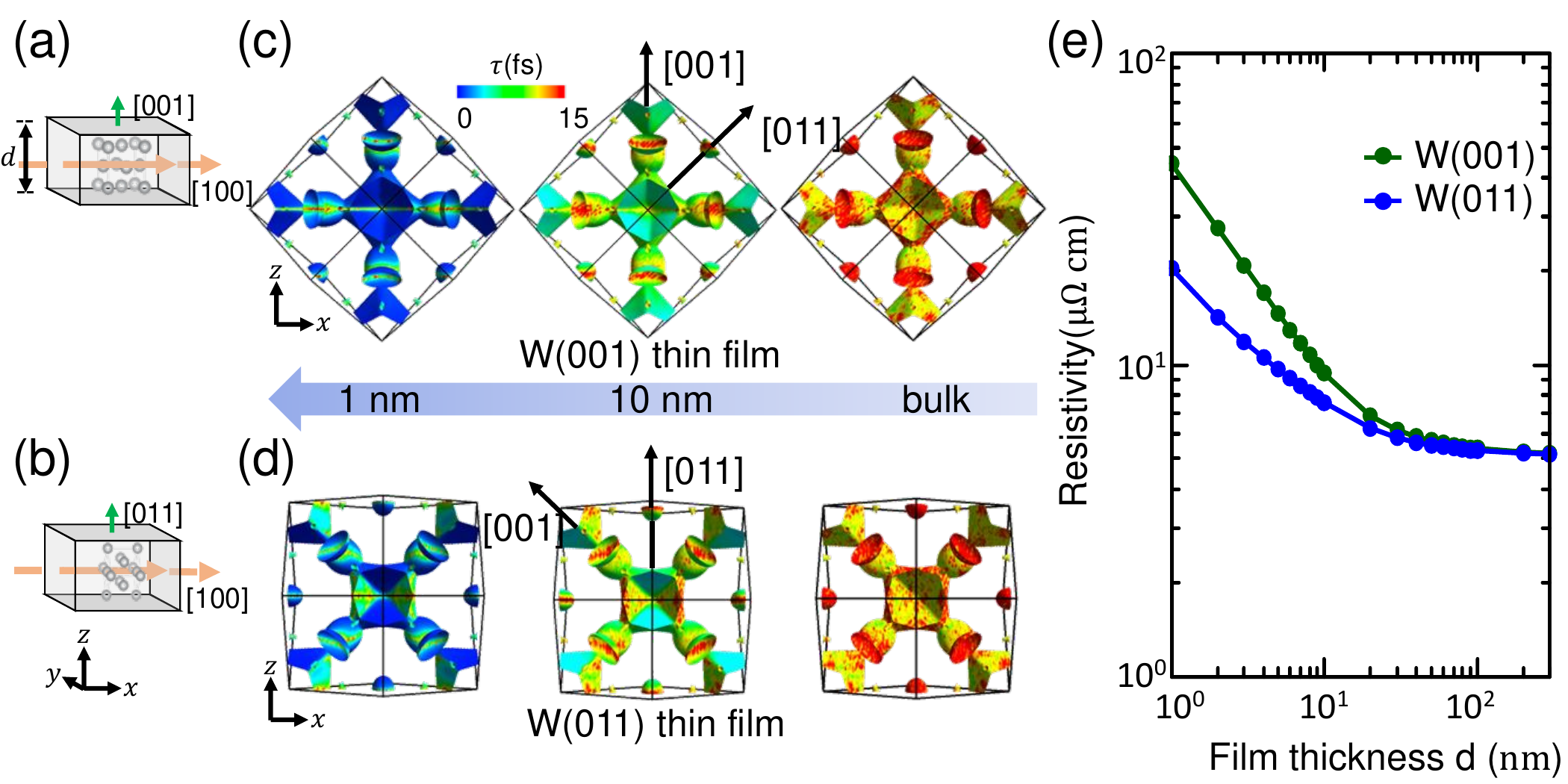}
\caption{\label{fig4} The thickness-dependent of resistivity for the anisotropic tungsten films. (a) Atomic structure of tungsten film oriented along the [001] direction within its conventional unit cell. (b) Atomic structure of a tungsten film oriented along the [011] direction. (c,d) Projected Fermi surfaces of tungsten films with (c) W(001) and (d) W(011) orientations, calculated using the modified relaxation-time model. Relaxation times are visualized on a consistent color scale for direct comparison across different film thicknesses: 1 nm, 10 nm, and bulk. (e) Calculated resistivity as a function of film thickness for W(001) and W(011) films, demonstrating pronounced anisotropic behavior driven by surface orientation.}
\end{figure*}
Building upon our validation of the momentum-dependent mean-free-path (MFP) model for isotropic metals, we extended our analysis to investigate the size-dependent resistivity of anisotropic materials, emphasizing how crystallographic orientation significantly impacts electron transport properties. Although isotropic metals such as Cu, Ag, Au, and Ir exhibit nearly spherical Fermi surfaces with direction-independent electronic behavior, many technologically relevant metals inherently exhibit pronounced anisotropy in their electronic structures, necessitating a comprehensive approach to understanding their properties.
To address this, we selected tungsten (W)—a prototypical anisotropic metal—due to its complex Fermi surface geometry and technologically important transport properties. Employing first-principles density functional theory (DFT) calculations coupled with our momentum-dependent MFP model, we explored tungsten films epitaxially grown along two distinct orientations, (001) and (011). 
The calculated bulk resistivity of tungsten at 300 K was found to be 4.33 $\mu\Omega\cdot\mathrm{cm}$, which is in reasonable agreement with the experimental value of 5.28$\mu\Omega\cdot\mathrm{cm}$~\cite{serway1998}, considering the typical discrepancies between DFT calculations and measured data.
This validation provides confidence in our model's robustness, which additionally enables precise evaluation of key transport parameters, such as the Fermi velocity, electron-phonon relaxation time, and mean-free-path distributions.
Our analysis, illustrated in Fig.~\ref{fig4}, demonstrates that although W(001) and W(011) differ only by a 45$^{\circ}$ crystallographic rotation, their electronic transport properties differ markedly. Specifically, this rotation leads to significant alterations in the projection of the bulk Fermi surface, and consequently, distinct momentum-dependent relaxation time distributions emerge for each orientation. For instance, in the W(001) films, electrons traveling perpendicular to the film plane have a longer bulk mean-free path ($\lambda_\mathrm{001}$), leading to a sharp resistivity increase at film thicknesses approaching this critical length scale. In contrast, W(011) films exhibit more gradual resistivity increases due to their inherently shorter bulk MFP ($\lambda_\mathrm{011}$).
Our model captures this critical orientation-dependent phenomenon by explicitly incorporating momentum directionality into the electron-phonon scattering calculations. In contrast, conventional isotropic models average over momentum directions, thereby eliminating the essential distinctions imposed by crystallographic orientation. As illustrated in Fig.~\ref{fig1}, isotropic averaging obscures the directional dependency, resulting in identical resistivity trends regardless of growth orientation.
Therefore, our results underscore the necessity and effectiveness of adopting a momentum-dependent MFP approach, particularly for materials with strong anisotropy. This advancement provides a more precise and physically accurate description of size-dependent resistivity in anisotropic metallic systems, offering valuable insights for their optimization in nanoscale electronic applications.

\subsection{Momentum-Dependent Mean-Free-Path Model Incorporating Grain Boundary Scattering in Thin Films of Anisotropic Metal}\label{subsec8}

\begin{figure}
\includegraphics[width=0.9\textwidth]{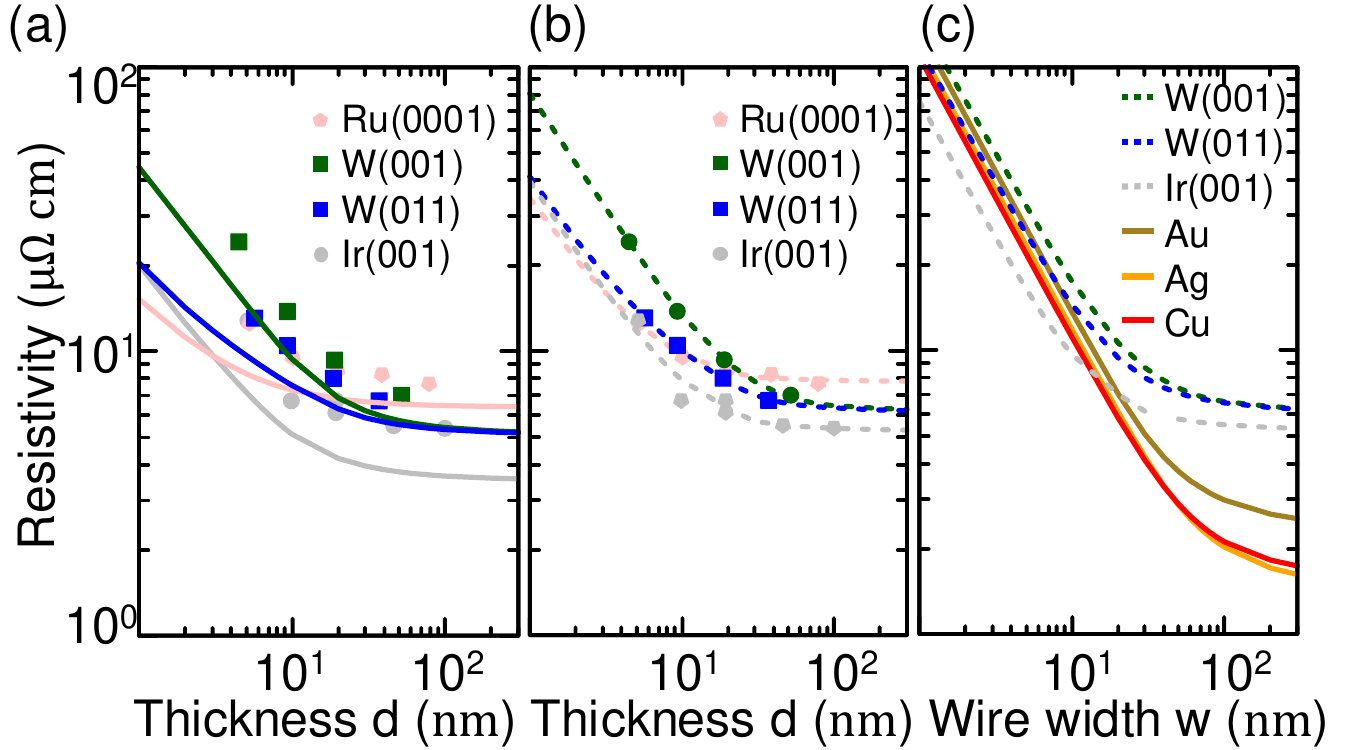}
\caption{The size-dependent resistivity effects driven by grain-boundary scattering in anisotropic metallic films and wires. (a) Size-dependent resistivity of anisotropic metallic thin films without grain boundary scattering. The analyzed materials include Ru(0001) (pink), W(001) (forest green), and W(011) (blue). Solid lines represent calculations obtained using the momentum-dependent mean-free-path (MFP) model, highlighting the influence of crystallographic orientation on in-plane electron transport. (b) Size-dependent resistivity of anisotropic metallic thin films incorporating grain boundary scattering effects. The same materials as in (a) are presented, with dashed lines indicating resistivity curves calculated by integrating grain boundary scattering via model fitting. Experimental data points correspond to previous literature values~\cite{Gall2020}. (c) Size-dependent resistivity of metallic square wires, comparing isotropic (Cu, Ag, Au, Ir) and anisotropic (W(001), W(011), Ru(0001)) materials. Solid lines represent resistivity calculations obtained from the momentum-dependent mean-free-path model. Directional notation for anisotropic materials indicates the surface orientation, with electron transport considered along the in-plane direction.}
\label{fig5}
\end{figure}
In the previous sections, we demonstrated that our momentum-dependent mean-free-path (MFP) model accurately predicts the size-dependent resistivity of both isotropic and anisotropic metals by explicitly considering crystallographic orientation effects. However, in practical thin-film applications, electron transport is strongly influenced by grain boundaries, necessitating an extension of our model to incorporate the effects of grain boundary scattering.
For noble metals such as Cu, Ag, and Au, the influence of grain boundary scattering on resistivity is generally minimal because of their intrinsically low grain boundary energies, which facilitate large grain growth. As illustrated in Fig.~\ref{fig3}(a), experimental bulk resistivity of these metals is closely aligned with theoretical predictions obtained from density functional theory (DFT) calculations considering only electron–phonon (el-ph) scattering. In contrast, materials with inherently higher grain boundary energies experience restricted grain growth, resulting in smaller grain sizes and significantly enhanced electron scattering at grain boundaries.
The tungsten (W) provides an instructive example: the bulk resistivity experimentally measured at 300 K is $5.3\times10^{-8}$ $\Omega\cdot\mathrm{m}$, which significantly exceeds our DFT-calculated resistivity ($4.33\times10^{-8}$ $\Omega\cdot\mathrm{m}$) based solely on electron–phonon (el-ph) scattering. The observed discrepancy highlights the need to incorporate additional scattering mechanisms, particularly grain boundary scattering, for accurate resistivity prediction.
We addressed this by implementing a grain boundary scattering model integrated into our momentum-dependent MFP framework. The key results of this analysis are illustrated in Fig.~\ref{fig5}. Fitting our model to experimental resistivity data yielded grain boundary sizes and scattering parameters: Ru(0001) exhibited a grain boundary size of 35.0 nm with grain boundary resistivity ($\gamma_{\mathrm{GB}}$) of $5.09\times10^{-16}$ $\Omega\cdot\mathrm{m}^{2}$; W(001) displayed 94.46 nm and $9.72\times10^{-16}$ $\Omega\cdot\mathrm{m}^{2}$, W(011) had 55.13 nm and $5.48\times10^{-16}$ $\Omega\cdot\mathrm{m}^{2}$, and Ir had 25.12 nm and $4.29\times10^{-16}$ $\Omega\cdot\mathrm{m}^{2}$. In particular, tungsten exhibited pronounced direction-dependent grain boundary scattering: The significant difference between the $\gamma_\mathrm{GB}$ values of W(001) and W(011) clearly indicates that the grain boundary properties depend on crystallographic orientation. This finding is consistent with previous studies reporting orientation-dependent grain boundary distributions driven by energetic considerations~\cite{chirayutthanasak2022}. In fact, low-temperature resistivity measurements of tungsten have also revealed a pronounced anisotropy, reporting a grain boundary resistivity difference of up to 2.7 times depending on the alignment of the grain boundary~\cite{Uray1991}. Such anisotropic behavior underscores the crucial importance of explicitly incorporating grain boundary scattering into resistivity models for anisotropic metallic films.
A key advantage of our approach lies in the precise separation of intrinsic electron-phonon scattering from grain-boundary scattering contributions. This capability allows us to isolate and quantify the grain boundary contribution, ultimately enabling more accurate predictions of resistivity in realistic thin-film and nanoscale materials. Such insights are crucial for optimizing thin-film synthesis and tuning of electrical transport properties through careful control of crystallographic orientation and grain boundary engineering.

\subsection{Momentum-Dependent Mean-Free-Path Model for Size-Dependent Resistivity in Wire Geometries}\label{subsec9}

Building upon our successful validation of the momentum-dependent mean-free-path (MFP) model for thin films, we further extended our analysis to metallic wires, which exhibit significantly enhanced surface scattering effects due to their geometry. Using a geometrically adapted version of our model, we calculated resistivity as a function of the width of the wire, revealing a pronounced increase compared to the thin film geometries. This sharper increase arises from the substantially elevated probability of electron-scattering events at wire surfaces, particularly as the wire dimension approaches the intrinsic bulk MFP scale.
Figure~\ref{fig5} (c) illustrates the resistivity trends for various metallic wires calculated using our momentum-dependent MFP model. In particular, wires exhibit a more pronounced resistivity enhancement compared to their thin-film counterparts, due to a higher likelihood of electron–surface interactions. This surface-dominated scattering mechanism becomes especially significant in narrower wires, highlighting the necessity of explicitly considering surface scattering effects to accurately predict electrical transport at nanoscale dimensions.
In the case of tungsten, we observe pronounced anisotropy-induced variations in resistivity trends for wires oriented along different crystallographic directions. W(001) and W(011) wires exhibit distinctly different resistivity behaviors, underscoring the critical role of crystallographic orientation in determining electron transport. Similarly to thin-film systems, wire resistivity increases sharply as their dimensions approach the bulk MFP scale, confirming the general correlation between the critical dimension and intrinsic electron transport lengths.
Furthermore, our model incorporates grain boundary scattering effects derived from experimental data, offering a comprehensive framework for realistic resistivity predictions. By explicitly including grain boundary contributions, we accurately quantify how resistivity is influenced by both surface scattering and internal microstructure, a feature particularly significant in materials such as tungsten, where anisotropic grain boundary scattering further modulates electron conduction. Consequently, tungsten wires exhibit pronounced differences in resistivity increases depending on their crystallographic orientation, highlighting the necessity of accounting for directional effects even within identical materials.
Collectively, our results demonstrate that the momentum-dependent MFP model effectively captures the complex interplay of surface and grain boundary scattering in wire geometries, providing a robust predictive tool for engineering optimal electronic performance in nanoscale conductors.

\subsection{Exploring Anisotropic MAX Phase Materials as Potential Alternatives to Copper in Ultra-Thin Interconnects: Application of Momentum-dependent MFP Model}\label{subsec10}

\begin{figure*}
\includegraphics[width=1.0\textwidth]{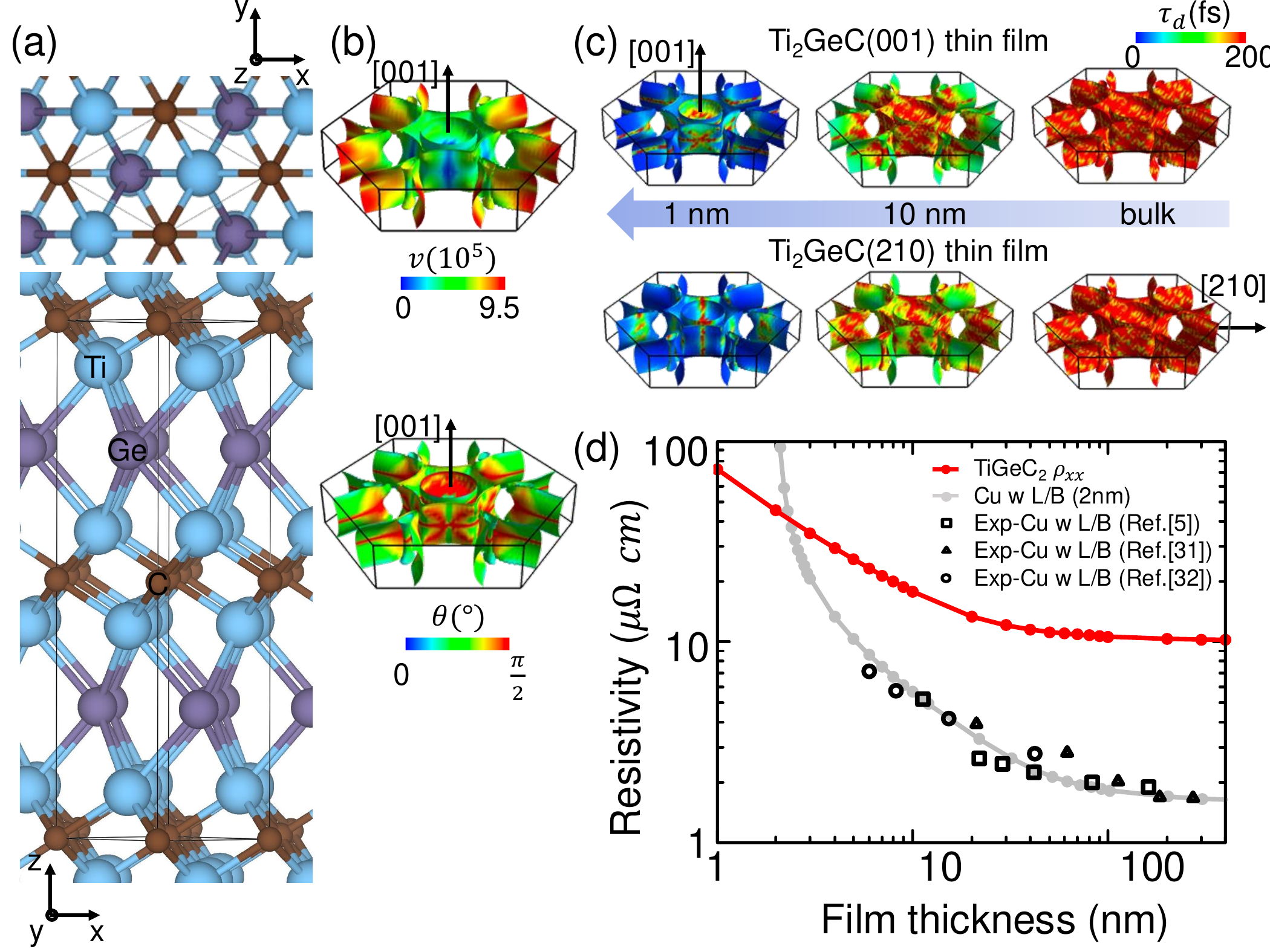}
\caption{Atomic structure and the thickness dependent of {Ti}{$_2$}GeC thin films with group velocity and relaxation time. (a) Crystal structure of {Ti}{$_2$}GeC, illustrating its layered atomic configuration and crystallographic orientations.
(b) The projected Fermi surface for the Fermi velocities and their angles for the relative surface.
(c) Projected the Fermi surface with relaxation time for {Ti}{$_2$}GeC thin films grown along the (100) and (001) orientations at various thickness of films. It illustrate how directional anisotropy affects electron transport characteristics. (d) Comparison of size-dependent resistivity at 300 K for isotropic Cu and anisotropic {Ti}{$_2$}GeC films. Experimental data (symbols) and model predictions (solid lines) are presented. Cu data includes films fabricated with interfacial liners to account for realistic boundary conditions. Experimental resistivity values for Cu are adapted from previous literature~\cite{Gall2020,Chawla2012,Khojier2013}. The results emphasize the critical role of anisotropy and crystallographic orientation in determining the size-dependent electrical transport properties.}
\label{fig6}
\end{figure*}

Given the increasing challenge of elevated resistivity in ultra-thin metallic conductors, identifying alternative interconnect materials superior to copper (Cu) has become essential. Our preceding analyses established the significant influence of anisotropy and crystallographic orientation on size-dependent resistivity. Leveraging these insights, we propose that anisotropic metallic systems could potentially outperform copper in ultra-thin interconnect applications.

Ideal interconnect materials should simultaneously fulfill two criteria: maintaining low bulk resistivity and minimizing the rate of resistivity increase as film thickness decreases. Achieving low bulk resistivity generally requires a long mean-free path (MFP) along the conduction direction. In contrast, minimizing size-induced resistivity enhancement requires shorter MFPs perpendicular to the conduction direction, reducing electron scattering at interfaces. This combination, however, is fundamentally incompatible in isotropic metals, as their uniform electronic structure inherently prevents directional tuning of the electron transport properties.

MAX phases - hexagonal layered carbides and nitrides described by the general formula $\mathrm{M_{n+1}AX_n}$ (where M represents early transition metals, A is from groups 13–14, and X denotes C or N) - present promising alternatives to traditional metals due to their strongly anisotropic quasi-two-dimensional electronic characteristics~\cite{Sankaran2021}. The layered structure of Ti$_2$GeC, a representative MAX-phase material, is illustrated in Fig.~\ref{fig6}(a). Its unique atomic stacking arrangement gives rise to strong anisotropy in electron transport.

This pronounced anisotropy is clearly visible in the angular dependence of the Fermi velocity and mean-free-path (MFP), as depicted in Fig.~\ref{fig6}(b). The quasi-two-dimensional nature of Ti$_2$GeC’s electronic structure significantly limits electron transport perpendicular to the layers, while facilitating efficient conduction within the layers.

Further applying our momentum-dependent MFP model to Ti$_2$GeC thin films with thicknesses down to 1 nm, we calculated modified relaxation time distributions for films oriented along the [210] and [001] directions, as presented in Fig.~\ref{fig6}(c). Notably, films grown along the (001) orientation exhibit longer preserved relaxation times as a result of their favorable alignment of electron transport paths relative to film boundaries, thus better retaining bulk-like conduction properties at reduced thickness.

Our DFT calculations predict a bulk resistivity of 10.12 $\mu\Omega\cdot\mathrm{cm}$ for Ti$_2$GeC at 300 K, approximately 6.37 times higher than that of Cu (1.59 $\mu\Omega\cdot\mathrm{cm}$). 
Although our theoretically estimated resistivity is slightly lower than the experimental values for Ti$_2$Ge at room temperature, which typically range between 15-20$\mu\Omega\cdot\mathrm{cm}$~\cite{Emmerlich2007,eklund2010}, this discrepancy is understandable, as our DFT-based calculation neglects grain boundary scattering and assumes an idealized bulk material.
While bulk Ti$_2$GeC resistivity is indeed higher than Cu, the size-dependent resistivity increase in Ti$_2$GeC is significantly slower due to its inherent anisotropy. Consequently, as the film thickness decreases, a resistivity crossover occurs near 2.5 nm, after which Ti$_2$GeC exhibits lower resistivity than copper. This effect is particularly significant given the known reliability issues of copper in nanoscale dimensions, including diffusion into dielectric layers, which severely degrade device performance~\cite{Chen2011,Li2020,bong2015,Caro2008,Mehta2017}. In contrast, Ti$_2$GeC, due to its robust anisotropic structure, effectively mitigates these surface-related limitations.

Therefore, although the bulk resistivity of Ti$_2$GeC (10.12 $\mu\Omega\cdot\mathrm{cm}$ from DFT) exceeds that of Cu, the reduction in size-induced resistivity increase presents a narrow-thickness regime where Ti$_2$GeC is superior for ultra-thin conductor applications. Our findings highlight the critical advantage of anisotropic materials and underline the importance of explicitly incorporating anisotropy and boundary scattering effects in resistivity modeling, opening new avenues for designing next-generation nanoscale interconnect materials.

\section{Conclusion}\label{sec5}
In this work, we developed a momentum-dependent mean-free-path (MFP) model to accurately capture size-dependent resistivity across isotropic and anisotropic metallic thin films. By integrating first-principles density functional theory (DFT) calculations with electron–phonon interactions, surface boundary scattering, and grain boundary scattering effects, we provided a comprehensive analysis of electron transport at nanoscale dimensions.
For isotropic noble metals (e.g., Cu, Ag, Au), our model closely reproduced conventional resistivity predictions because of their nearly spherical Fermi surfaces. However, anisotropic materials, exemplified by tungsten (W), exhibited pronounced directional resistivity variations, underscoring the necessity of explicitly incorporating crystallographic orientation in transport modeling. Additionally, we found significant direction-dependent grain boundary scattering effects in tungsten, which aligns closely with experimental observations and highlights the critical role of anisotropy in polycrystalline systems.
Motivated by emerging technological demands, we further explored MAX-phase materials, such as Ti$_2$GeC, as potential replacements for copper in ultra-thin interconnects. Despite their higher intrinsic resistivity, these anisotropic materials demonstrated reduced size-dependent resistivity increases compared to copper at nanoscale dimensions.
Our findings emphasize that explicitly incorporating anisotropy, grain boundary scattering, and directional transport effects significantly improves the accuracy of resistivity modeling. This momentum-dependent MFP model not only enhances predictive capabilities, but also informs the optimization and selection of novel materials to address resistivity challenges in next-generation ultra-thin electronic devices.

\nocite{*}
\bibliography{reference}

\end{document}